\documentclass[11pt]{article}
\usepackage{amsmath,amssymb,color,epsfig,cite}
\usepackage{graphicx}
\usepackage{subfigure}
\usepackage{setspace}

\usepackage{amsfonts}

\makeatletter
\@addtoreset{equation}{section}
\makeatother

\newcommand{\be}{\begin{equation}}
\newcommand{\ee}{\end{equation}}
\newcommand{\bea}{\setlength\arraycolsep{2pt} \begin{eqnarray}}
\newcommand{\eea}{\end{eqnarray}}
\newcommand{\nn}{\nonumber}

\def\ft#1#2{{\textstyle{\frac{\scriptstyle #1}{\scriptstyle #2} } }}
\def\fft#1#2{{\frac{#1}{#2}}}

\def\0{{\sst{(0)}}}
\def\1{{\sst{(1)}}}
\def\2{{\sst{(2)}}}
\def\3{{\sst{(3)}}}
\def\4{{\sst{(4)}}}
\def\5{{\sst{(5)}}}
\def\6{{\sst{(6)}}}
\def\7{{\sst{(7)}}}
\def\8{{\sst{(8)}}}
\def\sst#1{{\scriptscriptstyle #1}}

\begin{document}

\begin{flushright}

\end{flushright}

\begin{center}
{\large {\bf Violation of Thermal Conductivity Bound in Horndeski Theory }}

\vspace{20pt}
Hai-Shan Liu

\vspace{20pt}

{\it Institute for Advanced Physics \& Mathematics,\\
Zhejiang University of Technology, Hangzhou 310023, China}

\vspace{30pt}

\vspace{50pt}

\underline{ABSTRACT}

\end{center}

We consider charged AdS planar black holes in the four-dimensional Einstein-Maxwell-Horndeski theory with two free axions and analyse the black hole thermodynamics.  We calculate the holographic thermoelectric conductivities of the dual field theory and determine the ratio of thermal conductivity over the temperature. At low temperature and with zero electric current, we find that the ratio is proportional to temperature squared and hence can be arbitrarily small, providing the first example that violates the previously conjectured thermal conductivity bound.

\vfill {\footnotesize Emails: hsliu.zju@gmail.com }

\thispagestyle{empty}

\pagebreak




\section{Introduction}

The guage/gravity duality has brought many remarkable insights to the dynamics of some strongly coupled condensed matter systems \cite{ggd1,ggd2,ggd3,ggd4}. Several universal bounds of transport coefficients, including the well-known bound of the ratio of the shear viscosity to the entropy density \cite{KSS1,KSS2},  have been conjectured based on the holographic ``bottom-up" models. However, many of these bounds were later violated through various ways \cite{vioebd1,vioebd2}. One bound that stands out is that for the thermal conductivity $\kappa_{\sst{\rm DC}}$, at zero electric current.  This thermal bound was formulated by using a simple holographic model of Einstein-Maxwell-Dilaton theory, together with some momentum dissipation mechanism, which states that \cite{thebd}:
\bea
\fft {\kappa_{\sst{\rm DC}}}{ T} \ge {\cal C} \,, \label{thebd}
\eea
where ${\cal C}$ is a non-zero finite constant, as long as the dilaton potential is bounded from below \footnote{It is worth  pointing out that for unbounded potential the constant ${\cal C}$ approaches $1/(-V_{min})$, where $V_{min}$ is the minimum of the dilaton potential and approaches $- \infty$,   and thus $\kappa_{DC}/T$ can be zero at low temperatures as is mentioned in \cite{blake}.   }.  Remarkbly, this bound has been tested positive against a variety of holographic models, and up until now, there is no counterexample.

Motivated by searching for an counterexample, we study the holographic thermoelectric properties of black holes in Horndeski models. The Horndeski theories were constructed in 1970s \cite{hd1}, and they were rediscovered and have received much attention in cosmology \cite{hd2}.  A  particular  property of  Horndeski theories is that although their Lagrangians contain terms which have more than two derivatives,  the equations of motion involve at most two derivatives on each field.  This is analogous to Lovelock gravities \cite{ll}, and the theories can be ghost free.

Black hole solutions that are asymptotic to (locally) AdS spacetime have been constructed in Horndeski gravity theories in \cite{hs1, hs2} and the thermodynamics of these AdS black holes were analyzed in \cite{th1,th2}. The stability and causality were studied in \cite{stab1,stab2,stab3}. Holographic properties in Horndeski theories were deeply investigated in \cite{holo1,holo2,holo3,holo4,hdap4}. Further applications and properties were discussed in \cite{hdap1,hdap2,hdap3}. It turns out that for general coupling constants, there is no holographic $a$-theorem for Horndeski gravity; however, there exists a critical point, holographic $a$-theorem can be established \cite{hdap4}.  This implies that holographic application for Horndenski gravity is only sensible at the critical point of the coupling constant, where the dual field theory is scale invariant, rather than fully conformal invariant \cite{hdap4}.

In this paper, we shall consider $D=4$ Einstein-Maxwell-Horndeski gravity with a bare cosmological constant, together with two free axions for momentum dissipation. In section 2, we obtain charged AdS black holes at the critical point of the coupling constants. In section 3, we analyse the black hole thermodynamics. We study the holographic DC electrothermal conductivities in section 4 and find a counterexample of the bound (\ref{thebd}). We conclude the paper in section 5.

\section{The theory and AdS planar black holes}

The Lagrangian of Einstein-Maxwell-Horndeski theory with bare cosmological constant $\Lambda$ and two free axions $\phi_i$ is
\be
{\cal L } = \sqrt g \Big[ \kappa (R - 2 \Lambda - \ft 14 F^2)   - \ft12 (\alpha g^{\mu\nu} - \gamma G^{\mu\nu}) \partial_\mu \chi \partial_\nu \chi  - \ft 12 \sum_{i=1}^2 (\partial \phi_i)^2    \Big]\,,
\ee
where $F = dA$ and $G_{\mu\nu}$ is the Einstein tensor, and $\gamma$ is the Horndeski coupling.
The theory admits an AdS planar black hole solution \cite{holo1}:
\bea
ds^2 &=& - h(r) dt^2 + \fft{dr^2}{f(r)} + r^2 dx_i dx_i \,,\quad
 A = a(r) dt \,, \quad \chi = \chi(r) \,,\quad \phi_{1,2} = \lambda\, x_{1,2} \,,\nn\\
h &=& g^2 r^2 -\frac{2 \kappa  \lambda ^2}{\beta  \gamma +4 \kappa } -\frac{m_0}{r}
+\frac{\kappa  \left(3 g^2 q^2 (\beta  \gamma +4 \kappa )-\kappa  \lambda ^4\right)}{3 g^2 r^2 (\beta  \gamma +4 \kappa )^2}\nn\\
 &&-\frac{\kappa ^2 q^4}{60 g^2 r^6 (\beta  \gamma +4 \kappa )^2}-\frac{\kappa ^2 \lambda ^2 q^2}{9 g^2 r^4 (\beta  \gamma +4 \kappa )^2} \,, \cr
f &=& \frac{36 g^4 r^8 (\beta  \gamma +4 \kappa )^2}{\left(6 g^2 r^4 (\beta  \gamma +4 \kappa )-\kappa  (q^2 + 2  \lambda^2 r^2) \right)^2} \, h \,,\qquad
\chi' = \sqrt { \frac{6 \beta  \gamma  g^2 r^4- \kappa  (q^2 + 2  \lambda^2 r^2) }{6 \gamma  g^2 r^4} \, \fft 1 f} \,, \cr
a&=& a_0 -\frac{q}{r} +\frac{q \kappa  \lambda ^2}{9 g^2 r^3 (\beta  \gamma +4 \kappa )} +\frac{\kappa  q^3}{30 g^2 r^5 (\beta  \gamma +4 \kappa )}\,,
\eea
at the critical point of the coupling constants
\be
\Lambda = -\frac{3 g^2 (\beta  \gamma +2 \kappa )}{2 \kappa } \,, \quad \alpha = 3 \gamma  g^2 \,.
\label{crit}
\ee
The solution with $\lambda=0$ was constructed in \cite{hs2}. The parameters $a_0$, $q$, $m_0$ are integration constants.  When the parameters $(m_0,q,\lambda)$ all vanish, the solution becomes the planar AdS vacuum but with non-vanishing $\chi$, namely
\be
\chi=\chi_0+\sqrt{\beta} g^{-1} \log r\,.
\ee
Thus the special conformal transformation of the AdS vacuum is not preserved by $\chi$, but the Poincar\'e symmetry and the scale invariance survive, leading to a scale invariant quantum field theory \cite{hdap4}. It is important to note that at the critical condition (\ref{crit}) for the coupling constants, the Horndenski coupling $\gamma$ does not have zero limit; it is part of the vacuum construction analogous to the bare cosmological constant and should not be viewed as a perturbative parameter.  In fact for large AdS radius $\ell=1/g$ where the AdS/CFT correspondence is applicable, the critical condition (\ref{crit}) implies that the Horndeski coupling constant must be large, which is significant different from the Lovelock theories, or the non-critical cases where $\gamma$ is fixed and can be arbitrarily small.

There is a curvature singularity located at $r=r_*$, where $f$ diverges, and it is determined by
\be
F(r_*) \equiv 6 g^2 r_*^4 (\beta  \gamma +4 \kappa )-\kappa  (q^2 + 2  \lambda^2 r_*^2)  = 0 \,. \label{sing}
\ee
In order to describe a black hole, we require that the largest root $r_*$ should be inside the event horizon, that  is  $r_* < r_0$, where the radius of the event horizon $r_0$ is the largest root of $f(r_0) =0$. The Hawking temperature can be obtained through standard method
\be
T = \frac{6 g^2 r_0^4 (\beta  \gamma +4 \kappa )-\kappa  q^2 -2r_0^2  \kappa  \lambda ^2}{8 \pi  r_0^3 (\beta  \gamma +4 \kappa )} \,.\label{temp}
\ee
It is worthwhile pointing out that the requirement $r_* < r_0$ guarantees that the temperature is positive definite. In particular, the temperature can be arbitrarily close to zero, but cannot reach zero.

Though the linearized equations of motion for Horndeski theory
involve only two
derivatives, it is still necessary to insure that the kinetic
term of the Horndeski scalar is positive to avoid any possible ghostlike behavior.  The kinetic term for the axion perturbation $\delta \chi$  is $P^{00}\,\delta \dot\chi\, \delta\dot \chi$, where
\be
P^{00} = \frac{144 \gamma  g^4 \kappa  r^6 (\beta  \gamma +4 \kappa )^2 \left(q^2+\lambda ^2 r^2\right)}{F(r)^3} \,.\label{P00}
\ee
Here $F(r)$ is defined in (\ref{sing}). In order to avoid any ghostlike excitation, $P^{00}$
should be non-negative from horizon to asymptotic infinity. As was discussed above, the requirement $r_* < r_0$ implies that $F(r)$ is always positive on and outside the horizon. It follows that the positivity of $P^{00}$ requires that $\gamma > 0$, which is consistent with the critical condition (\ref{crit}).

\section{Black hole thermodynamics}

The thermodynamics of an AdS black hole in Horndeski theory have been studied extensively in \cite{th1,th2} with the aid of Wald formalism \cite{wald1,wald2}. Here, we begin by reviewing the main results.
The variation of Lagrangian ${\cal L}$ gives the equations of motion and total derivative terms
\be
\delta {\cal L} = \text{e.o.m} + \sqrt g \nabla_\mu J^\mu \,,
\ee
from which we can define a $1$- form $J_\1 = J_\mu dx^\mu$ and its Hodge dual $\Theta_{\3} = -1 * J_\1$. Specializing the variation to be induced by an infinitesimal diffeomorphism $\delta x^\mu = \xi^\mu$, we can define a $3$-form and show that
\be
J_{(3)} \equiv \Theta_{(3)} - i_\xi * {\cal L} = \text{e.o.m} - d * J_\2 \,,
\ee
where $i_\xi$ represents a contraction of $\xi^\mu$ with the 3-form $* {\cal L}$  and $J_\2 = d \xi$. Then, one can define a 2-form $Q_\2 \equiv * J_\2$, such that $J_\3 = d Q_\2$ on shell. The variation of the Hamiltonian is given by
\bea
\delta H=\delta Q -i_\xi \Theta &=& -  2 r \sqrt{\frac{h}{f}} \Big(\kappa  + \fft{\gamma}{4} f \chi'^2  \Big ) \delta f  \, \Omega_{(2)} \cr
& & -  r^{2}  \sqrt{\frac{h}{f}}  \kappa\,  \Big(\fft{f}{h}\,
a\delta a' + \fft{a a'}{2} (\fft{\delta f}{h} -
\fft{f \delta h}{h^2})\Big) \, \Omega_{(2)}\,.
\eea
Note that here we consider $\lambda$ and $g=1/\ell$ as thermodynamical constants. We also choose a gauge that the electric potential vanishes on the horizon, then the variation of the Hamiltonian on the horizon is
\be
\delta H_+ = 16 \pi T ( \kappa + \fft{\gamma}{4} f \chi'^2 ) \delta (\fft{r_0^2}{4}) \,,
\ee
where $T$ is given in (\ref{temp}), while at the infinity, it is
\be
\delta H_\infty = \kappa \mu \delta q - \fft{(4 \kappa + \beta \gamma)\delta m_0}{2} \,,
\ee
And Wald showed that the variation of the Hamiltonian vanishes on the Cauchy surface. For a black hole it implies that $\delta H_+ + \delta H_\infty = 0$, which gives the first law of the thermodynamics,
\be
d M = T d S + \Phi_e d Q_e + \Phi_\chi^+ d Q^+_\chi\,. \label{1stlaw}
\ee
with
\bea
M &=& \ft{1}{2} (4 \kappa + \beta  \gamma  )  m_0 \,, \qquad \Phi_e = \mu \,, \qquad Q_e = \kappa q  \,, \cr
S &=& \Big(\kappa + \fft {\gamma} {4} (f \chi'^2)|_{r_0} \Big) 4 \pi r_0^2 = \frac{16 \pi  r_0 (\beta  \gamma +4 \kappa )}{3 g^2} \, T\,,\cr
\Phi_\chi^+ &=& -\fft{\gamma\, r_0^2 \, T }{8}\sqrt{f \chi'^2|_{r_0}} \,,\cr
Q_\chi^+ &=&16 \pi \int_{r=r_0}  \sqrt{(\partial\chi)_{+}^2}= 16 \pi  \sqrt{f}\,\chi'\Big|_{r=r_0}\,.
\eea
$\Phi_\chi^+ $ and $ Q^+_\chi$ are the scalar potential and charge respectively, (see \cite{th2} for more details). The temperature is given in (\ref{temp}). It is important to note that owing to our parametrization we appear to be able to set $\gamma=0$ in the above thermodynamical quantities; however, as we have remarked earlier, there is no smooth limit of $\gamma=0$.

\section{Holographic thermoelectric conductivity}

There are various ways to obtain the holographic DC thermoelectric conductivities \cite{DChong,DC1,DC2,DC3}. The key step is to construct the relevant radially conserved current, which serves as a bridge connecting the boundary physical properties to the black hole horizon information.
We consider the following perturbations around the background solution,
\bea
\delta g_{tx_1} = t U_1(r) + \Psi_{tx_1} \,, \quad \delta g_{rx_1} =  \Psi_{rx_1} \,, \quad \delta A_{x_1} = t U_2(r) + a_{x_1} \,, \quad \delta \phi_1 = \fft{\Phi(r)}{\lambda} \,.
\eea
The radially conserved electric current can be easily obtained with the help of Maxwell equation $\partial_r ( \sqrt g F^{rx_1}) = 0$,
\be
{\cal J} = \kappa \sqrt g F^{rx_1} \,.
\ee
The radially conserved holographic heat current is more difficult to construct, since its conservation involves both Einstein and Maxwell equations. Fortunately, a general formula of deriving the holographic heat current was proposed in \cite{holo4} by using Noether symmetry for general classes of gravity theories. Applying this formula in our theory, we obtain the holographic heat current
\be
{\cal Q} = \sqrt g \Big( \kappa( 2 \nabla^r \xi^{x_1} + a F^{rx_1})  + \fft \gamma 2 \, g^{rr} (\partial_r \chi)^2  \nabla^r \xi^{x_1}  \Big) \,,
\ee
where $\xi$ is the time-like Killing vector $\partial_t$. We find that the electric and heat current can be time independent by choosing
\be
U_1 = - \zeta h \,, \qquad U_2 = - E + \zeta a \,,
\ee
where $E$ and $\zeta$ are constants which parameterize the sources for the electric and heat currents, respectively. Near the black hole horizon, we impose the ingoing wave condition
\be
a_{x_1}' = \fft{-E + \zeta a}{\sqrt {hf}}  + \dots \,, \qquad \Psi_{tx_1} = \Psi^{(0)}_{tx_1} - \zeta h \int \fft{1}{\sqrt{hf}}  + \dots \,,
\ee
where  $\Psi^{(0)}_{tx_1}$ is a regular function whose value on the horizon can be determined by the linearized perturbative equation of motion
\be
\Psi^{(0)}_{tx_1} (r_0) = -\frac{48 E g^2 \kappa  q r_0^5 (\beta  \gamma +4 \kappa )+\zeta  \left(\kappa  \left(q^2+2 \lambda ^2 r_0^2\right)-6 g^2 r^4 (\beta  \gamma +4 \kappa )\right)^2}{48 g^2 \kappa  \lambda ^2 r_0^5 (\beta  \gamma +4 \kappa )} \,.
\ee
Now, we are in a position to evaluate the radially conserved currents on the horizon
\bea
{\cal J} &=& \big( \kappa +\frac{\kappa  q^2}{\lambda ^2 r_0^2} \big) E + \frac{4 \pi ^2 q (\beta  \gamma +4 \kappa )}{3 g^2 \lambda ^2 r_0} \, T^2 \, \zeta \,, \cr
{\cal Q} &=&  \frac{4 \pi ^2 q (\beta  \gamma +4 \kappa )}{3 g^2 \lambda ^2 r} \, T^2\, E     + \frac{16 \pi ^4 (\beta  \gamma +4 \kappa )^2}{9 g^4 \kappa  \lambda ^2} \, T^4 \, \zeta \,.
\eea
The DC conductivity matrix is then given by
\bea
\sigma_{\sst{\rm DC}} &=& \fft{\partial {\cal J}}{\partial E} =   \kappa( 1  + \frac{  q^2 }{ r_0^2 \lambda^2  } ) \,, \qquad\qquad\quad
\alpha_{\sst{\rm DC}} = \fft 1 T  \fft{\partial {\cal J}}{\partial \zeta} = \frac{4 \pi ^2   q (\beta  \gamma +4 \kappa )}{ 3 g^2 r_0 \lambda^2} \, T\,, \cr
\bar \alpha_{\sst{\rm DC}} &=& \fft 1 T  \fft{\partial {\cal Q}}{\partial E} =  \frac{4 \pi ^2 q (\beta  \gamma +4 \kappa )}{ 3 g^2 r_0 \lambda^2 } \, T\,, \qquad
\bar \kappa_{\sst{\rm DC}} = \fft 1 T  \fft{\partial {\cal Q}}{\partial \zeta} =  \fft{16 \pi ^4 (\beta  \gamma +4 \kappa )^2}{9 \kappa g^4 \lambda^2} \, T^3 \,.
\eea
It is thus clear that the electric bound, which was proposed in \cite{elebd}, is satisfied
\be
\sigma_{\sst{\rm DC}} = \kappa( 1  + \frac{ q^2 }{ r_0^2 \lambda^2 } ) \ge 1 \,. \label{elebd}
\ee

The form of the electric conductivity $\sigma_{\sst{\rm DC}}$ is the same as that of Einstein-Maxwell case \cite{EMDC} while it is expressed in terms of black hole horizon radius $r_0$, as is pointed out in \cite{holo1}. It is easy to check that $\alpha_{\sst{\rm DC}} = \bar \alpha_{\sst{\rm DC}}$, which means the Onsager relation holds. Furthermore, we also find that the thermal relation $
S T \alpha_{\sst{\rm DC}} - Q_e \bar \kappa_{\sst{\rm DC}} = 0 $
holds for this system.

 The thermal conductivity at zero electric current is
\be
\kappa_{\sst{\rm DC}} = \frac{16 \pi ^4 (\beta  \gamma +4 \kappa )^2}{9 \kappa g^4 \left(\lambda^2+\frac{  q^2}{r_0^2}\right)} \, T^3 \,. \label{thercd}
\ee
There are two more quantities of interest, the Lorentz ratios of the thermal conductivities over the electric conductivities, which are given by
\be
\bar L = \fft{\bar \kappa_{\sst{\rm DC}}}{\sigma_{\sst{\rm DC}} T} =  \frac{S^2}{\kappa ^2 \left(q^2+\lambda ^2 r_0^2\right)}  \,,\qquad
L = \fft{\kappa_{\sst{\rm DC}}}{\sigma_{\sst{\rm DC}} T} =  \frac{\lambda ^2 r_0^2 S^2}{\kappa ^2 \left(q^2+\lambda ^2 r_0^2\right)^2} \,.\label{lr}
\ee
Usually, the Lorentz ratio $L$ is a constant, due to the fact that the heat transport and the electric transport both involve the charge carriers, like free electrons in metal, which is well known as the Wiedemann-Franz law. As we can see, this law is violated in our case, which may be explained in terms of independent transportation of charge and heat in a strongly coupled system.

It was observed that there is a bound for Lorentz ratio $\bar L$ in \cite{DC3}
\be
\bar L \le \fft{S^2}{Q_e^2} \,.
\ee
From (\ref{lr}), we can see this bound is indeed satisfied in our case
\be
\bar L  =  \frac{S^2}{\kappa ^2 \left(q^2+\lambda ^2 r_0^2\right)}  \le  \fft{S^2}{Q_e^2}  \,. \label{lrbd}
\ee

Having established that both the electric bound (\ref{elebd}) and the Lorentz ratio bound (\ref{lrbd}) are satisfied, we now examine the thermal conductivity bound (\ref{thebd}). It follows from (\ref{thercd}), we find
\be
\fft {\kappa_{\sst{\rm DC}}}{ T }= \frac{16 \pi ^4 (\beta  \gamma +4 \kappa )^2}{9 \kappa g^4 \left(\lambda^2  +\frac{  q^2}{r_0^2}\right)} \, T^2  \,.
\ee
As discussed in section 2, although the temperature of the black hole cannot be zero, it can be arbitrarily close to zero. For sufficiently low temperature, the ratio becomes
\be
\fft{\kappa_{\sst{\rm DC}}}{T} \sim \frac{16 \pi ^4 (\beta  \gamma +4 \kappa )^2}{9 g^4 \sqrt{\kappa } \sqrt{6 g^2 q^2 (\beta  \gamma +4 \kappa )+\kappa  \lambda ^4}} \, T^2 + {\cal O} (T^3) \,.
\ee
Thus, the ratio can be arbitrarily small at low temperature and it is obvious that the scalar potential of Horndeski theory is bounded, hence our holographic model violates the proposed thermal conductivity bound (\ref{thebd}).

\section{Conclusions}

In this paper, we considered the Einstein-Maxwell-Hordeski theory with bare cosmological constant and two free axions. The theory admits analytical charged AdS planar black holes, where the axions span over the two-dimensional plane. It is important to note that the AdS spacetime where the black holes are immersed in is the vacuum solution at the critical point of the coupling constants.  The special conformal transformation of the AdS is broken by the Horndeski scalar $\chi$, but the Poincar\'e and scale invariance survive, giving rise to scale invariant quantum field theory at the boundary. We analysed the thermodynamics of the black hole and calculated the holographic thermoelectric conductivities of the dual field theory. The focus of the paper is to examine various related universal holographic bounds proposed in literature.

The Horndeski term doesn't contribute directly to the electric conductivity, which takes the same form as that of Einstein-Maxwell theory. Thus the electric conductivity bound is preserved. Furthermore, the Onsager relation $\alpha_{\sst{\rm DC}} = \bar \alpha_{\sst{\rm DC}}$ and the thermal relation $ST \alpha_{\sst{\rm DC}} - Q_e \bar \kappa_{\sst{\rm DC}} = 0$ are both satisfied. The situation for the thermal conductivity, on the other hand, is quite different. Although the Lorentz ratio bound is satisfied, the ratio of thermal conductivity at zero electric current over temperature turns out to be, at low temperature, constantly proportional to the square of the temperature; therefore, it can be arbitrarily small as the temperature is low, violating the conductivity bound (\ref{thebd}). This rare counterexample indicates that an underlying principal is needed to understand the condition when the bound is valid.  It is of interest to investigate whether the breaking of the conformal symmetry to sale invariance is the culprit for the bound violation.

\section*{Acknowledgement}

We are grateful to anonymous referees for their useful suggestions which improve the paper a lot. H-S.L.~is supported in part by NSFC grants No.~11305140, No.~11375153,
No.~11475148 and No.~11675144.

\end{document}